# Group Handover Management in Mobile Femtocellular Network Deployment

Mostafa Zaman Chowdhury, Sung Hun Chae, and Yeong Min Jang[*]
Department of Electronics Engineering, Kookmin University, Seoul 136-702, Korea
E-mail: mzceee@yahoo.com, hooni129@hanmail.net, yjang@kookmin.ac.kr

**Abstract:** The mobile femtocell is the new paradigm for the femtocellular network deployment. It can enhance the service quality for the users inside the vehicles. The deployment of mobile femtocells generates lot of handover calls. Also, number of group handover scenarios are found in mobile femtocellular network deployment. In this paper, we focus on the resource management for the group handover in mobile femtocellular network deployment. We discuss a number of group handover scenarios. We propose a resource management scheme that contains bandwidth adaptation policy and dynamic bandwidth reservation policy. The simulation results show that the proposed bandwidth management scheme significantly reduces the handover call dropping probability without reducing the bandwidth utilization.

## I. Introduction

The femto-access-points (FAPs) are low-power, small-size home-placed base stations (BSs) that enhance the service quality for the indoor mobile users [1]-[4]. Currently the mobile users inside the vehicular environments suffer from several difficulties e.g., low *SNIR* level, higher outage probability, and lower throughput due to the poor signal quality inside the vehicle. Femtocells deployment in the vehicular environments, we refer as the mobile femtocell can solve these difficulties [2]. In mobile femtocellular network deployment, the mobile station (MS) is connected to the indoor FAP instead of outside macrocellular or the satellite networks. Therefore, the MS can receive better quality signal. Table 1 shows some differences between the mobile femtocell and the fixed femtocell. The backhauling networks are the wireless links for the mobile femtocellular networks that is the most important difference between the fixed femtocellular networks and the mobile femtocellular networks deployments. Fig. 1 shows the device-to-CN (core network) connectivity for the mobile femtocellular networks deployment. Except the backhauling networks connectivity, the other network entities are same as the fixed femtocellular networks deployment.

There are several issues those have to be solved to deploy femtocellular networks in vehicular environments. The backhaul network is the wireless link. This is the most challenging issue for the mobile femtocellular network deployment. Appropriate backhaul network selection should be considered for designing of mobile femtocellular network. The wireless backhaul networks also cause the handover between the backhaul networks. The interference management can be done by proper and efficiently allocating the frequency among different femtocells. The handover management is another issue for the mobile femtocellular network deployment. Individual handovers as well as the group handovers are happened. Sufficient amount of resources should be provided to mobile femtocellular networks during the backhaul network handovers, group handovers, and individual handovers. Therefore, we focus on the resource management for the group handover in mobile femtocellular network deployment. We discuss a number of group handover scenarios. We propose a resource management scheme that contains bandwidth adaptation policy and dynamic bandwidth reservation policy. The bandwidth adaptation [5] policy permits the releasing of some bandwidth from the quality of service (QoS) adaptive multimedia traffic calls to accommodate handover calls in the system. The dynamic bandwidth reservation is based on the backhaul network handover and mobile femtocell-to-macrocell handover rates. Using the simulation results, we show that the proposed bandwidth management scheme significantly reduces the handover call dropping probability without reducing the bandwidth utilization.

**Table 1.** Comparison between the fixed FAP and mobile FAP

|  | **Fixed Femtocell** | **Mobile femtocell** |
|---|---|---|
| **Location** | Home, office, and other places where FAP location is fixed | Bus, train, car, and other vehicular environments where the location of FAP is continuously changed due to the movement of the vehicle. |
| **Interference** | From neighbor FAPs located in the nearby apartment, office, and etc. | From neighbor FAPs located in other vehicles. |
| **Role of FAP** | Main access point for the user | Works as a realy station |
| **Backhaul network for femtocell traffic** | Normally xDSL, CATV, and others broadband wired networks | Satelitte, Mobile WiMAX, LTE-Advanced, or other broadband wireless networks |
| **Backhaul network selection** | No | There is backhaul network selection whenever more than one networks are available |
| **Relay functionality** | No | The combination of the outside antenna and FAPs works as a relay station |

---
[*]Corresponding author of this paper

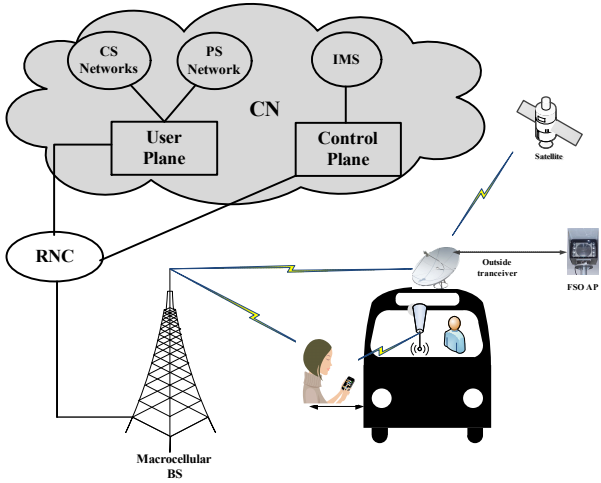

**Fig. 1.** Network architecture for the femtocellular network deployment in vehicular environment.

The rest of this paper is organized as follows. Group handover concept for the mobile femtocellular cases is presented in Section II. The scenarios of group handovers are also discussed in this section. In Section III, we provide the proposed resource allocation policy. Performance evaluation results of the proposed schemes are presented and compared in Section IV. Finally, Section V concludes our work.

## II. Group Handover

The ability to seamlessly switch between the femtocells and the macrocell networks is a key driver for femtocellular network deployment. Also, handover between the backhaul networks is a major concern for the mobile femtocellular network deployment. In case of mobile femtocellular network deployment, a group of users may change the access networks simultaneously or within short time. Group handover may be happened due to the movement of individual users or due to the movement of the femtocellular networks. Therefore, the resources of the macrocellular networks should be carefully handled to accommodate these handover calls.

When a train or a bus or any vehicle containing the mobile femtocellular network arrives at station, a number of mobile users get off from the vehicle and a number of users enter into the vehicle. Hence, the group handover concept is applied for the mobile femtocellular network deployment. For the mobile femtocellular network deployment, the individual user does not need to be handed over to the target macrocellular BS if the vehicle is in movement. In this case, only the outside transceiver performs the handover between the backhaul networks. However, for this case, the target macrocellular BS have to provide sufficient resources to accommodate the mobile femtocellular network. The individual user does not need to make handover between the macrocellular BSs.

The group handover is also happened when the bus/train arrives at the station and a number of active users get off from the bus/train. Fig. 2 (a) shows the group handover scenario under the same macrocellular BS in mobile femtocellular network deployment. In this case, the present backhaul network and the target macrocellular BS for the user is same. Fig. 2 (b) shows the group handover scenario under the other macrocellular BS in mobile femtocellular network deployment. In this case, the present macrocellular BS and the target macrocellular BS for the user are not same. For both the cases, huge number of handover is occurred when the vehicle reaches to the station.

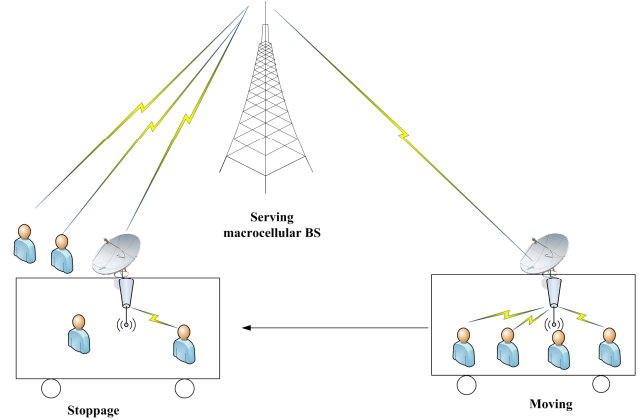

(a).

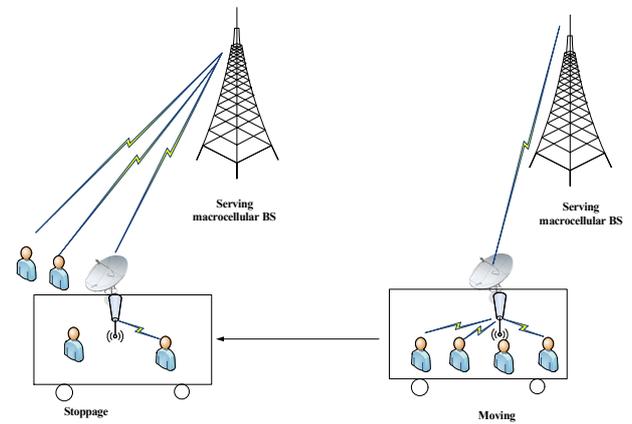

(b)

**Fig. 2.** Group handover scenario (a) under the same macrocellular BS in mobile femtocellular network deployment (b) under the two separate macrocellular BS in mobile femtocellular network deployment.

## III. Resource Allocation in Macrocellular Networks to Support Group Handover

Radio resource allocation in macrocellular networks for group handover in mobile femtocellular network deployment is important. Because, due to the mobile femtocellular network deployment, the femtocell-to-macrocell handover rate is increased. Also a group of users connecting to a mobile femtocell moves from one macrocell coverage to another macrocell coverage. Therefore, macrocell system needs to accommodate the movement of mobile femtocells. Otherwise all the users will lose their connections with the femtocell. To support theses handover calls, efficient radio resource management scheme is required. Fig. 3 shows the proposed radio resource allocation in macrocellular networks for group handovers in mobile femtocellular network deployment. $C_{vacant}$, $C_{occupied}$, and $C_{releasable}$ are, respectively, the vacant bandwidth due to macrocell-to-femtocell handover and due to the movement of mobile femtocellular networks from one macrocell to another macrocell during time $T$, the occupied bandwidth by the existing calls, and the amount of releasable bandwidth from the existing non-real-time calls. Therefore, the priority of the handover calls in the macrocellular networks is given by few reserved bandwidth and the QoS adaptation.

### A. QoS Adaptation

Our scheme allows the adaptation of QoS adaptive calls. The system can retrieve $C_{releasable}$ amount of bandwidth from the running QoS adaptive calls to accommodate the handover calls. The QoS adaptability [5] of some multimedia traffic calls allows the reclaiming of system bandwidth to support more number of handover calls. The releasable amount of bandwidth from QoS adaptive call is calculated as:

$$\beta_{releasable(i)} = \xi_i \beta_{r,i} \quad (1)$$

$$C_{releasable} = \sum_{i=1}^{M} N_i \xi_i \beta_{r,i} \quad (2)$$

where $\xi_i$ is the maximum portion of bandwidth that can be degraded for a background traffic call of $i$-th class to accept a handover call in the system. $\beta_{r,i}$ is the requested bandwidth for a background traffic call of class $i$. $M$ is the total number of existing QoS adaptive traffic class. $N_i$ is the total number of existing calls of traffic class $i$.

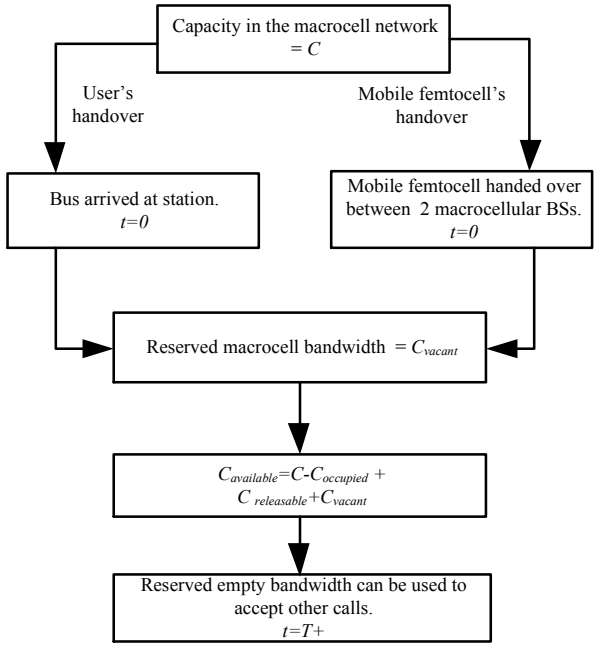

**Fig. 3.** Radio resource allocation in macrocellular networks for group handover in mobile femtocellular network deployment.

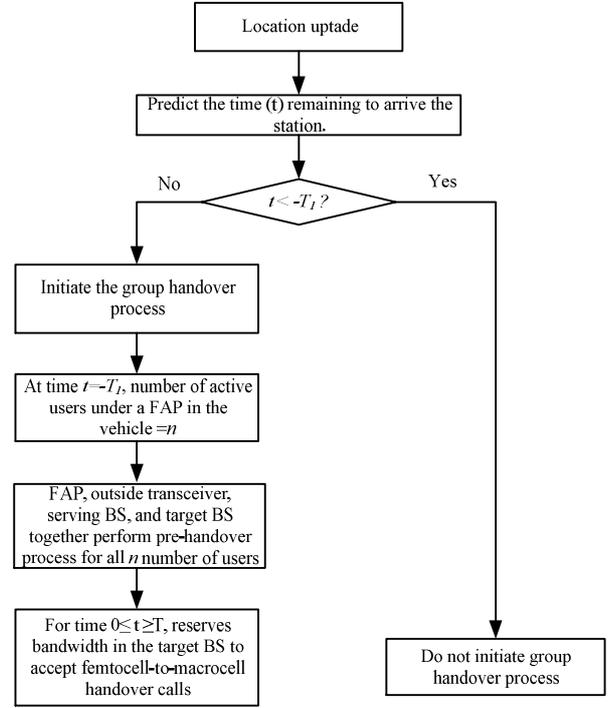

**Fig. 4.** Steps of bandwidth reservation for user's level handover in mobile femtocellular network deployment.

### B. Bandwidth Reservation

*Case 1 (user's movement between femtocell and macrocell):* In this case, freed bandwidth due to the movement of the users from the macrocell-to-femtocell handover is reserved for the time $T$ to accommodate handover calls. This phenomenon is occurred for the scenario of Fig. 2.

*Case 2 (mobile femtocell moves from one macrocell to another macrocell):* When a mobile femtocell moves from one macrocell to another macrocell coverage area, the freed bandwidth from the previous macrocellular network will be reserved for the threshold time $T$ to accommodate handover calls.

Therefore, dynamically variable $C_{vacant}$ amount of bandwidth is reserved to accommodate the handover calls. The key feature of this bandwidth reservation is that, the reserved amount of bandwidth is increased if the traffic due to the mobile femtocells is increased.

Fig. 4 shows the steps of bandwidth reservation for user's level handover in mobile femtocellular network deployment. $T_1$ time before reaching the station, the system starts the bandwidth reservation and group handover process. All the $n$ number of active users in between time $t=-T_1$ and $t=0$ are considered for the group handover process.

## IV. Performance Analysis

In this section, we evaluate the performances of the proposed scheme. Table 2 summarizes the values of the parameters that we used in our analysis. We consider the macrocellular networks as the backhauling networks.

Table 2: Summary of the parameter values used in analysis

| Parameter | Value |
|---|---|
| Bandwidth capacity of macrocellular networks | 6 Mbps |
| Maximum portion of bandwidth that can be degraded for a QoS adaptive traffic call of *i-th* class ($\xi_i$) | 0.5 |
| Ratio of QoS adaptive traffic and non-QoS adaptive traffic | 1 : 1 |
| Macrocell –to-macrocell handover call arrival rate : mobile femtocell-to-macrocell handover call arrival rate | 1 : 1 |
| Average call duration time considering all calls (exponentially distributed) | 120 sec |
| Average cell dwell time for the macrocell users (exponentially distributed) | 540 sec |
| Threshold time for bandwidth reservation (*T*) | 10 sec |

Fig. 5 shows that the proposed scheme provides negligible handover call dropping probability even for very high traffic condition. The other schemes cannot provide lower handover call dropping probability due to the non-priority of handover calls. Fig. 6 shows the bandwidth utilization comparison. Even though our scheme contains bandwidth reservation and the priority of the handover calls, the proposed scheme does not significantly reduce the bandwidth utilization.

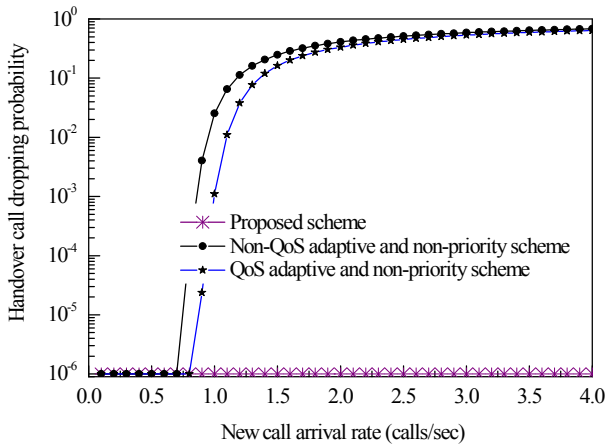

**Fig. 5**. Comparison of handover call dropping probability.

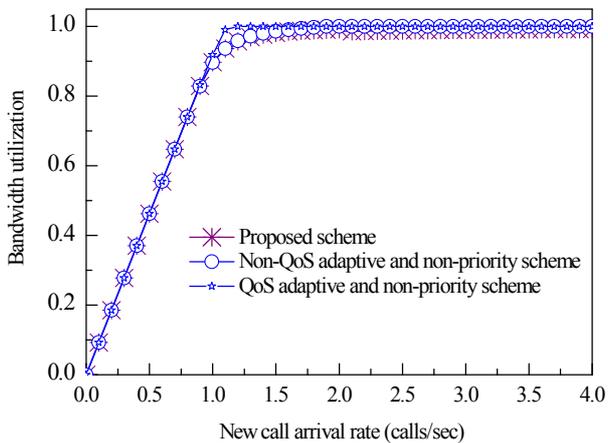

**Fig. 6.** Comparison of bandwidth utilization.

## V. Conclusions

The femtocells in the vehicular environment i.e., the mobile femtocells will be the new paradigm of the femtocellular network deployment. However, the deployment of mobile femtocells will create huge handover calls with the macrocellular networks. The group handovers need to be managed carefully. Otherwise huge number of running calls will be dropped. The proposed scheme is able to efficiently manage the bandwidth of the macrocellular networks to handle huge number of handover calls.

In this paper, we discussed the group handover issues in mobile femtocellular network deployment. Various scenarios of group handover are shown. The simulation results demonstrate that the proposed scheme can significantly reduce the handover call dropping probability without reducing the bandwidth utilization.

## Acknowledgments

This work was supported by the IT R&D program of MKE/KEIT [10035362, Development of Home Network Technology based on LED-ID].